\documentstyle[aps,prb,epsf,multicol]{revtex}
\begin{document}
\draft
\title{Flux-noise spectra around the Kosterlitz-Thouless
transition \\ for two-dimensional superconductors}
\author {Beom Jun Kim and Petter Minnhagen}
\address {Department of Theoretical Physics,
Ume{\aa} University,
901 87 Ume{\aa}, Sweden}
\preprint{\today}
\maketitle
\begin{abstract}

The flux-noise spectra around the Kosterlitz-Thouless transition
are obtained from simulations of the two-dimensional resistively shunted
junction model. In particular the dependence on the distance $d$ between
the pick-up coil and the sample is investigated. The typical experimental situation
corresponds to the large-$d$ limit and a simple relation valid in this
limit between the complex impedance and the noise spectra is clarified.
Features, which distinguish  between the large- and small-$d$ limit, are 
identified and the possibility of observing these features in experiments is discussed.
\end{abstract}

\pacs{PACS numbers: 74.40+k, 74.76.-w, 74.50+r}


\begin{multicols}{2}

\section{Introduction}\label{sec_intro}

Spontaneously created vortices drive the transition between the
superconducting and normal state for thin-film superconductors and
two-dimensional (2D) Josephson-Junction arrays
(JJA's).~\cite{minnhagen-rev} This means that the physics of the
vortices is responsible for the characteristic features in a region
around the transition. One manifestation of this is the static
characteristics of the phase transition which is of the
Kosterlitz-Thouless (KT) type.~\cite{minnhagen-rev,kt} Another manifestation is the
dynamical features around the transition. These are
reflected in the flux-noise spectra and the complex impedance.
In the present paper we investigate the flux-noise spectra through computer simulations of the  
resistively shunted Josephson junction (RSJ) model on a 2D square lattice. In
particular we clarify the relation between the flux-noise spectra and
the complex impedance. 

There have been a number of recent  experimental~\cite{ferrari,rogers,shaw,uppsala} as well as
theoretical studies~\cite{houlrik,hwang,tiesinga,wagenblast,timm,capezzali} dealing with the flux-noise
spectra. The typical experimental setup measures the
fluctuation of the magnetic flux through a pick-up coil situated at a
distance above the sample.~\cite{ferrari,rogers,shaw,uppsala} Many simulation studies on the
other hand have measured the vorticity fluctuation associated with a fixed
area of the sample itself.~\cite{hwang,tiesinga,wagenblast} It has been assumed that this would
roughly correspond to the magnetic flux spectra of the
experiments. However, in the present investigation we show that there
are significant differences. The typical experimental situation
corresponds to the limit of large distance between the pick-up coil
and the sample. In this limit there exists a simple
relation~\cite{houlrik} between the flux-noise spectra and the complex
impedance of the sample, which we here verify both directly from the simulations and through
analytic calculations.

In Sec.~\ref{sec_formalism} we describe how the flux-noise spectrum is obtained from
simulations of the RSJ model. Section~\ref{sec_theory} clarifies the relation
between the flux-noise spectrum and the complex impedance. The results
from the simulations are described and discussed in Sec.~\ref{sec_results}.
Particular attention is given to the implication for experimental measurements
of flux noise and the complex impedance. Finally Sec.~\ref{sec_conc} contains
some concluding remarks.

\section{RSJ model and Flux Noise} \label{sec_formalism}

\subsection{RSJ model and numerical method}
\label{sub_model} 
In our simulations we use the 2D RSJ model on a  square lattice 
with periodic boundary conditions (PBC). This is usually assumed to be 
a good model of a Josephson junction array (JJA). In particular, from  
the point of view of 
vortex fluctuations and vortex dynamics, it is expected to have 
the same physics as a thin superconducting 
film, as well as a JJA.~\cite{minnhagen-rev} There might however be some differences in 
the level of vortex fluctuations around the transition.~\cite{jonssona}

The RSJ model incorporates the condition of the
local current conservation and the equations of motion can be written as:~\cite{minnhagen-rev}
\begin{equation} \label{eq_motion} 
\dot\theta_i = -\sum_j G_{ij} {\sum_k}^{'}(\sin\phi_{jk} + \eta_{jk}), 
\end{equation} 
where $G_{ij}$ is the square lattice Green function,  
the primed summation is over the four nearest neighbors of the site $j$, and   
$\phi_{jk} \equiv \theta_j - \theta_k$ with the phase $\theta_j$ of the complex 
order parameter  at site $j$. Here we measure time  $t$
in units of $\hbar/2eRI_c$, where $R$ is the shunt resistance and $I_c$ 
is the critical current of a single junction. 
The thermal noise current $\eta_{jk}$ in units of $I_c$  
satisfies the conditions $\langle \eta_{ij}(t) \rangle = 0$ and 
\begin{equation}  
\langle \eta_{ij}(t) \eta_{kl} (0)\rangle = 
2T(\delta_{ik}\delta_{jl} -  \delta_{il}\delta_{jk})\delta(t) , 
\end{equation} 
where $\langle \cdots \rangle$ is the ensemble average, 
and the temperature $T$ is in units of the Josephson coupling strength 
$J \equiv \hbar I_c/2e$.   

The RSJ model may be used to calculate the current distribution in
the limit of a large perpendicular penetration length
$\Lambda=\Phi_0c/4\pi^2I_c$, where $\Phi_0$ is the flux quantum;
This is typically the case for a 2D superconductor.~\cite{minnhagen-rev} 
In this limit one may replace the gauge
invariant phase difference $\phi_{jk} \equiv \theta_j - \theta_k
-A_{jk}$, where
$A_{jk}\equiv \frac{2\pi}{\Phi_0}\int_j^k{\bf A}({\bf
    r})\cdot d{\bf l}$ is the line integral of the vector potential,
  with $\phi_{jk} \equiv \theta_j - \theta_k$ because the coupling to
  the electromagnetic self-field is in this limit so weak that it has
 little influence on the fluctuations of the supercurrent (we are here considering 
the situation without an external magnetic field).~\cite{minnhagen-rev}

We measure the flux-noise spectrum $S(\omega)$ defined by the Fourier transformation: 
\begin{equation} \label{eq_Sw} 
S(\omega) = \int_{-\infty }^{\infty }dt  e^{i\omega t} S(t),  
\end{equation} 
where $S(t)$ is the time-correlation function of the magnetic flux 
$\Phi(t)$ through a pick-up coil: 
\begin{equation} \label{eq_St} 
S(t) = \langle \Phi(t) \Phi(0) \rangle  
\end{equation} 
(see Sec.~\ref{sub_dipole} for the calculation of $\Phi$). 
In addition, we calculate the dynamic dielectric function 
$1/\epsilon(\omega)$ of the vortices in the Coulomb-gas analogy 
given by~\cite{jonsson} 
\begin{eqnarray} \label{eq_Re_eps} 
{\rm Re}\left[\frac{1}{\epsilon(\omega)}\right] &=& 
        \frac{1}{\epsilon(0)}+\frac{2\pi\omega T^{\rm CG}}{T^2} 
      \int_0^\infty dt \sin \omega t \, G(t) ,  \label{reps} \\ 
{\rm Im}\left[ \frac{1}{\epsilon(\omega)}\right] &=& 
-\frac{2\pi\omega T^{\rm CG}}{T^2}\int_0^\infty dt \cos \omega t \, G(t) , \label{eq_Im_eps} 
\end{eqnarray} 
where $T^{\rm CG} = T/(2\pi J \langle \cos\phi) \rangle$ is the Coulomb-gas temperature, 
and the time-correlation function $G(t)$ is defined by  
\begin{eqnarray} 
G(t) &\equiv& \frac{1}{L^2}\langle F(t) F(0) \rangle , \label{eq_Gt} \\ 
F(t) &\equiv& \sum_{\langle ij \rangle_x} \sin\phi_{ij} , 
\end{eqnarray}  
and the sum is over all links in $x$-direction. 
 
The dynamic dielectric constant $1/\epsilon(\omega)$ is related to the 
conductivity $\sigma(\omega)$ and the complex impedance $Z(\omega)$ by~\cite{beom} 
\begin{equation} 
\sigma(\omega)=Z^{-1}(\omega)=1-\frac{T}{i 2\pi \omega T^{CG}}\frac{1}{\epsilon(\omega)}  .
\label{sigma} 
\end{equation} 
  
In the numerical simulation of an $L \times L$ array (in most cases $L=64$ 
and occasionally $L=128$ are used in  
the present paper), we use the periodic boundary condition for the phase variables,  
i.e., $\theta_{i + L \hat{\bf x}} = \theta_{i + L \hat{\bf y}} = \theta_i$, and   
the thermal noise currents are generated from the uniform probability distribution. 
For the time integration of equations of motion in Eq.~(\ref{eq_motion}), we use 
the Euler method with the discrete time step $\Delta t = 0.05$. 
In practice we have calculated $S(t)$ up to a $t_{\rm max}$ beyond which $S(t)$
became so small that the simulations could not be converged well enough to obtain
further information. In our simulations this turned out to be $t_{\rm max} \approx 100$
for $T > 1.10$ and $t_{\rm max} \approx 400$ for lower temperatures. This means that
we could not reach frequencies below $\omega \lesssim 0.016$ directly from the 
simulation data. However, when presenting the data we have for convenience
used an extrapolation to large $t$ based on an expected large-$t$ behavior
(this extrapolation does not change the behavior
in frequency range $\omega \gtrsim 0.016$).
From the ergodicity 
of the system, we can change the ensemble averages of the form $\langle O(t) O(0)\rangle$ 
in Eqs.~(\ref{eq_St}) and (\ref{eq_Gt}) 
to the average $\langle O(t+t') O(t') \rangle_{t'}$ over time $t'$, 
and averages over more than $10^7$ time steps were typically performed.

\subsection{Flux noise and the dipole approximation} \label{sub_dipole} 
As mentioned above, previous numerical 
simulations~\cite{hwang,tiesinga} usually calculate the noise spectra 
from the fluctuation of the {\em vorticity} defined by the directional 
sum of the gauge-invariant phase difference around each plaquette. The 
fluctuations of the total vorticity over an area of the sample is then 
used to estimate the flux-noise spectrum. 
On the other hand in experiments~\cite{ferrari,rogers,shaw,uppsala} 
the fluctuations of  
the {\em magnetic flux} penetrating a pick-up coil situated at a distance 
above the sample is measured.
The relation between the vorticity noise  
and the flux noise has so far not been studied in detail and is the 
subject of the present paper. 
 
We consider three distinct cases: 
The first one is the fluctuation of the magnetic flux associated with the vortices on a 
fixed area $A$ of the sample. The vortices describe the rotation of the 
supercurrent on the sample.~\cite{minnhagen-rev} This means that 
the magnetic flux associated with an elementary plaquette is
proportional to the 
rotation of the supercurrent around the plaquette. For the 
RSJ model the magnetic flux for a plaquette is then given by~\cite{houlrik,jonsson,olsson} 
 \label{eq_n}
\begin{equation} 
n({\bf r}) \equiv \frac{T^{\rm CG}}{T} \sum_p \sin\phi_{ij},  
\label{eq_nn}
\end{equation} 
in units of the flux quantum $\phi_0$.~\cite{foot-nr} 
Here ${\bf r}$ is the central position of the plaquette and 
the summation is around the plaquette $p$. Consequently the  
total magnetic flux associated with the rotation of the 
supercurrent at a given time $t$ is 
\[ \Phi(t)=\int_A d^2r \; n({\bf r},t) ,
\] 
where the integral over position denotes the sum over all elementary 
squares inside the area $A$. 
 
The second case is the fluctuation of the vorticity associated with a 
fixed area $A$ of the sample. The vorticity of an elementary plaquette 
is given by~\cite{minnhagen-rev} 
\begin{equation} \label{eq_v1} 
v({\bf r}) \equiv \frac{1}{2\pi} \sum_{p} \phi_{ij}, 
\end{equation} 
where the phase difference $\phi_{ij}$ is restricted to the interval $-\pi<\phi_{ij}\leq\pi$.  
The total vorticity $V(t)$ associated with the area $A$ is consequently 
\[ V(t)=\int_A d^2 r \; v({\bf r},t)  .
\] 
 
The third case  
corresponds to the experimental situation where one measures the magnetic flux 
through a pick-up coil situated a distance $d$ from the sample.
Note that the first case corresponds to the case when $d=0$. 
However, in typical experiments $d$ is a macroscopic length (typically 
of the order of 0.1 mm).~\cite{rogers,shaw,uppsala} 
 
\begin{figure}
\centerline{\epsfxsize=7cm \epsfbox{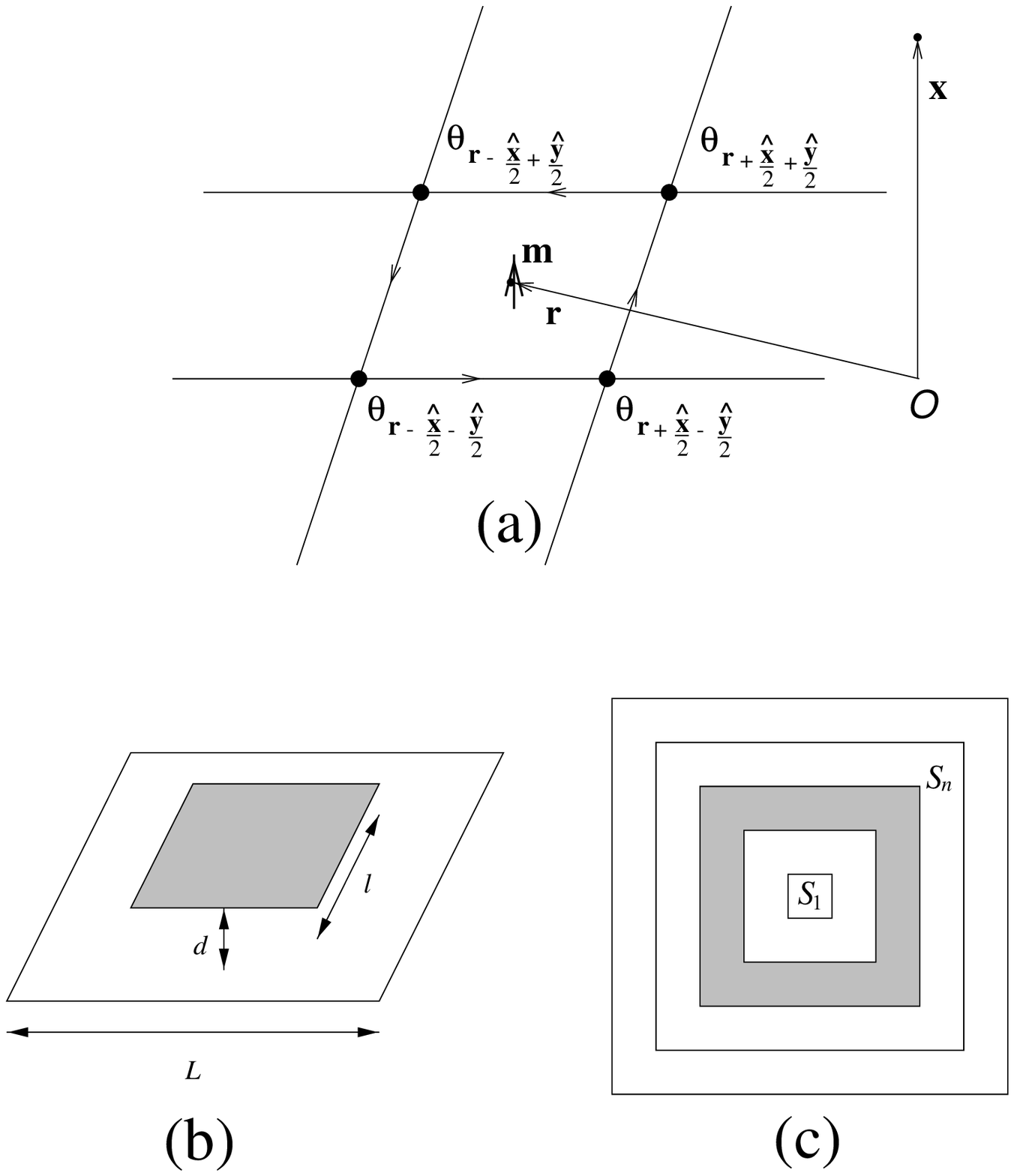}}
\vskip 1cm
\caption{(a) The vortex dipole moment associated with the dual lattice
point ${\bf r}$ is estimated by the circulating current around the
plaquette.
The magnetic field at the observation point ${\bf x}$ is the summation of the 
contributions from all such vortex dipole moments.
(b) The $l \times l$ pick-up coil is separated by the distance $d$ from
the $L \times L$ square array of Josephson junctions. 
(c) The whole array is divided into quadratic enclosures. The
plaquettes surrounding such an enclosure is denoted by $S_n$ (shaded area).  
The magnetic flux due the plaquettes belonging to $S_n$ is calculated.}
\label{fig_geom}
\end{figure}

In this section, we focus on the third case and use a dipole approximation
to obtain an expression for the magnetic flux through a pick-up coil at a
distance $d$ from the sample where $d\gg \Lambda$. At this distance the
magnetic field distribution from a vortex is of dipole form. In the
continuum limit only the circulation of the supercurrent around a closed
loop which encloses a vortex core region gives a finite value. This value
can be positive or negative but has the same magnitude for all closed loops
which enclose vortex cores.~\cite{minnhagen-rev} This means that one can estimate the magnetic
field at $d\gg\Lambda$ from the circulation of the supercurrent around
small closed loops which cover the area of the superconductor by
associating each of these loops with a dipole field where the strength of
the dipole moment is proportional to the circulation of the supercurrent.
This approximation is readily carried over to the array.  The circulation
of the supercurrent around a plaquette in units of $I_c$ is given by [see
Fig.~\ref{fig_geom}(a)]~\cite{foot-m}
\end{multicols} 
\noindent\rule{0.5\textwidth}{0.1ex}\rule{0.1ex}{2ex}\hfill 
\begin{eqnarray} 
 & &  \sin\left[\phi\left({\bf r}-\frac{\hat {\bf x}}{2}-\frac{\hat {\bf y}}{2}, 
      {\bf r}+\frac{\hat {\bf x}}{2}-\frac{\hat {\bf y}}{2}\right)\right]  
   +\sin\left[\phi\left({\bf r}+\frac{\hat {\bf x}}{2}-\frac{\hat {\bf y}}{2}, 
      {\bf r}+\frac{\hat {\bf x}}{2}+\frac{\hat {\bf y}}{2}\right)\right]  \nonumber \\ 
 & & + \sin\left[\phi\left({\bf r}+\frac{\hat {\bf x}}{2}+\frac{\hat {\bf y}}{2}, 
      {\bf r}-\frac{\hat {\bf x}}{2}+\frac{\hat {\bf y}}{2}\right)\right]  
   + \sin\left[\phi\left({\bf r}-\frac{\hat {\bf x}}{2}+\frac{\hat {\bf y}}{2}, 
      {\bf r}-\frac{\hat {\bf x}}{2}-\frac{\hat {\bf y}}{2}\right)\right] ,  
\label{eq_rotj}
\end{eqnarray}
where $\phi({\bf r'}, {\bf r''}) \equiv \theta_{\bf r'} - \theta_{\bf r''}$ and 
$\theta_{\bf r'}$ is the phase  of the complex order parameter at site ${\bf r'}$.  
The contribution to the magnetic field at large distances from this
circulation can then approximately be described in terms of a dipole moment
${\bf m_r}=m_{\bf r}\hat{\bf  z}$ proportional to the circulation.
The magnetic field at the observation point ${\bf x}$ due to this
vortex dipole moment ${\bf m}_{\bf r}$ is given by~\cite{jackson} 
\begin{equation} 
{\bf B}({\bf x}) = \sum_{\bf r} \frac{ 3{\bf e}_{\bf rx} ( {\bf e}_{\bf rx} 
\cdot {\bf m}_{\bf r}) - {\bf m}_{\bf r} }{ | {\bf x} - {\bf r} |^3},  
\end{equation} 
where ${\bf e}_{\bf rx} \equiv ( {\bf x} - {\bf r} )/|{\bf x} - {\bf 
  r}|$ is the unit vector in the direction of ${\bf x} - {\bf r}$ and 
the summation is over all dual lattice points (central positions of plaquettes).  
The magnetic flux through the pick-up coil [the shaded area 
in Fig.~\ref{fig_geom}(b)] separated by distance $d$ from the array is given by the 
surface integral: 
\begin{equation} 
\Phi = \int_{\rm coil} {\bf B}({\bf x})\cdot d{\bf s}  . 
\label{Phi} 
\end{equation} 
We consider the case $d\gg a$ so that the magnetic field inside the pick-up coil does not 
vary much on a microscopic length scale $a$. We may then replace 
the integral Eq.~(\ref{Phi}) by the discrete summation: 
\begin{eqnarray} \label{eq_full} 
\Phi &\approx &\sum_{ {\bf r'} \in l\times l}  
    {\bf B}({\bf r'} + d\hat{\bf z})\cdot \hat{\bf z} \nonumber \\ 
&=& \sum_{ {\bf r'} \in l\times l} \sum_{{\bf r} \in L\times L} 
 \frac{ 3{\bf n}_{\bf rr'} ( {\bf n}_{\bf rr'} 
\cdot {\bf m}_{\bf r}) - {\bf m}_{\bf r} }{ | {\bf r'} + d\hat{\bf z} - {\bf r} |^3} \cdot 
\hat{\bf z} \nonumber \\ 
&=& \sum_{ {\bf r'} \in l\times l} \sum_{{\bf r} \in L\times L} 
 \frac{ 3d^2 - | {\bf r'} + d\hat{\bf z} - {\bf r} |^2 } { | {\bf r'} + d\hat{\bf z} - {\bf r} |^5} 
m_{\bf r} , 
\end{eqnarray} 
where ${\bf x}$ is decomposed into ${\bf x}={\bf r'} + d\hat{\bf z}$, and the two 2D vectors   
  ${\bf r}$ and ${\bf r'}$ denote positions on  
the array. The summations $\sum_{\bf r'}$ and $\sum_{\bf r}$ are performed 
on the $l \times l$ pick-up coil and the $L \times L$ whole array, respectively [see Fig.~\ref{fig_geom}(b)]. 
Since Eq.~(\ref{eq_full}) contains $O(L^4)$ terms (in this work we choose $l = L/2$),  
the calculation of the magnetic flux in this way  
requires most of the computer time in the numerical simulations. This time-consuming part is avoided in the 
approximate scheme we use in our simulation to obtain the results described in Sec.~\ref{sec_results}. 
 
\begin{figure}
\centerline{\epsfxsize=8cm \epsfbox{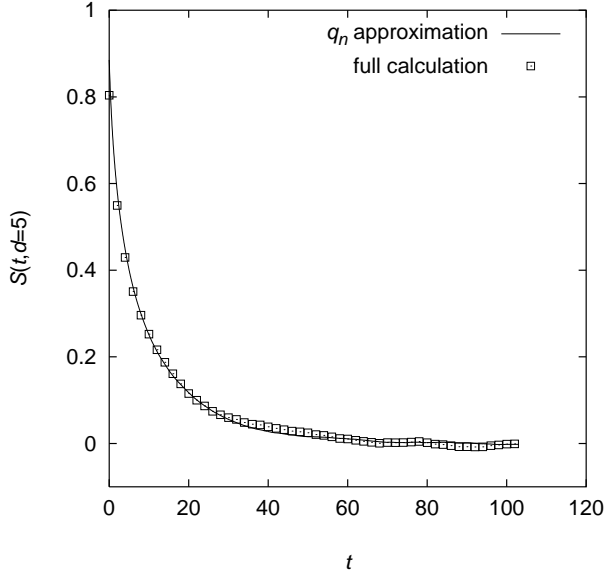}}
\vskip 0.5cm
\caption{Comparison between the flux noise from the full calculation
[Eq.~(\ref{eq_full})] and from the approximate scheme 
[Eqs.~(\protect\ref{eq_Phi}), (\protect\ref{eq_Vn}), and
(\protect\ref{eq_qn})]. The figure shows the flux noise as a function of time $t$,
$S(t,d) \equiv\langle \Phi(t,d) \Phi(0,d)\rangle$ for a  $16 \times 16$ coil with the distance
$d=5$ from a $32\times 32$ array at the temperature
$T=1.10$ (in units of $J/k_B$). As seen the approximation scheme
reproduces the full calculation very accurately.} 
\label{fig_full}
\end{figure}

We use the following approximate scheme: 
The whole array is divided into stripes formed by elementary plaquettes which 
enclose the midpoint of the pick-up coil as the sides of a square.   
This is illustrated in Fig.~\ref{fig_geom}(c). Each such collection of
 elementary plaquettes are denoted by $S_n$ where $2n-1$ is the number of plaquettes 
of each of the four stripes forming the sides of the square.  
The magnetic flux in Eq.~(\ref{eq_full}) is expressed as the summation of the contributions from each $S_n$: 
\begin{equation} \label{eq_Phi} 
\Phi = \sum_n q_n M_n, 
\end{equation} 
where $M_n$ is the summation 
of the vortex dipole moments for the plaquettes forming $S_n$ and $q_n$ is the appropriate weight factor: 
\begin{eqnarray} 
q_n &\equiv& \left(\sum_{ {\bf r} \in S_n} m_{\bf r} \sum_{ {\bf r'} \in l\times l} 
\frac{ 3d^2 - | {\bf r'} + d\hat{\bf z} - {\bf r} |^2 }  
{ | {\bf r'} + d\hat{\bf z} - {\bf r} |^5}\right)
 \left( \frac{1}{\sum_{ {\bf r} \in S_n} m_{\bf r}}\right), \label{eq_qn1} \\ 
M_n &\equiv& \sum_{ {\bf r} \in S_n} m_{\bf r} \label{eq_Vn}. 
\end{eqnarray} 
\hfill\raisebox{-1.9ex}{\rule{0.1ex}{2ex}}\rule{0.5\textwidth}{0.1ex} 
\begin{multicols}{2} 
\noindent
We now assume that $m_{\bf r}$ in Eq.~(\ref{eq_qn1}) can to good
approximation
 be replaced by the average value for each $S_n$,  i.e., 
 $ \tilde{m}_{\bf r}=\sum_{ {\bf r} \in S_n} m_{\bf r}/A_n$ where
$A_n$ is the number of plaquettes contained in $S_n$.   
This simplifies Eq.~(\ref{eq_qn1}) to: 
\begin{equation}\label{eq_qn} 
q_n = \frac{1}{A_n} \sum_{ {\bf r} \in S_n} \sum_{ {\bf r'} \in l\times l} 
\frac{ 3d^2 - | {\bf r'} + d\hat{\bf z} - {\bf r} |^2 }  
{ | {\bf r'} + d\hat{\bf z} - {\bf r} |^5}.  
\end{equation} 
The magnetic flux in Eq.~(\ref{eq_Phi}) can by aid of Eqs.~(\ref{eq_Vn}) 
and (\ref{eq_qn}) be computed in $O(L^2)$  
operations, since $q_n$ within this approximation is a purely geometric  
quantity which is independent of time.  
In Fig.~\ref{fig_full} we compare the flux noise  
$S(t,d) \equiv \langle \Phi(t,d) \Phi(0,d) \rangle$ from the full calculation in  
Eq.~(\ref{eq_full}) and the $q_n$-approximation in Eqs.~(\ref{eq_Phi}) and (\ref{eq_qn}) 
for a $32 \times 32$ array with a $16 \times 16$ coil size and the distance $d=5$.  
It is clearly shown that the approximation made in Eq.~(\ref{eq_qn}) is indeed a very good approximation. 
 
\begin{figure}
\centerline{\epsfxsize=8cm \epsfbox{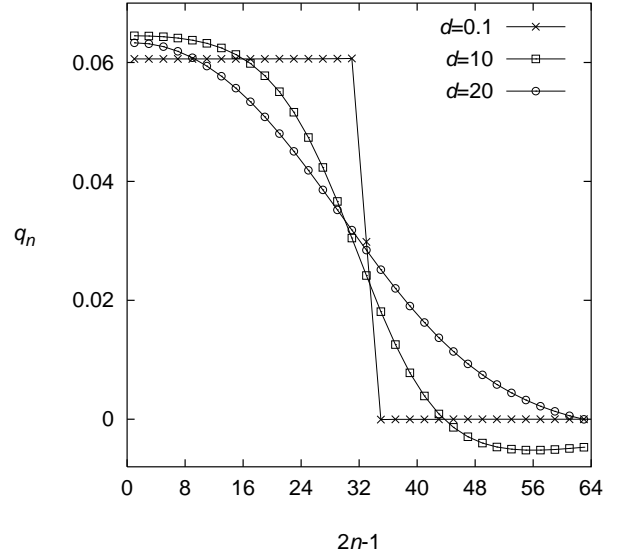}}
\vskip 0.5cm
\caption{The weight factor $q_n$ (in arbitrary unit) 
for $S_n$ plotted against the size $2n-1$ (in numbers of
plaquettes per side).
The data are for a $64 \times 64$ array with the coil size $32 \times
32$ for different distances $d$. When the 
distance $d$ between the coil and the array is very small, only plaquettes 
inside of the pick-up coil contribute to the magnetic flux,
whereas all plaquettes contribute for larger $d$.
}
\label{fig_qn}
\end{figure}

Figure~\ref{fig_qn} shows $q_n$ as a function of the linear size ($2n-1$) 
of $S_n$ for $d = 0.1, 10$, and 20 (a $32 \times 32$ pick-up coil 
is used for a $64\times 64$ array). For very small values of $d$, $q_n$ becomes 
a step function where only $S_n$'s inside the pick-up coil contribute 
to the magnetic flux. On the other hand, as $d$ is increased, it is clearly 
seen that there is a significant contribution to the magnetic field caused by the $S_n$'s  
outside  
the pick-up coil area. In previous numerical studies of the flux-noise spectra 
the magnetic flux has usually been approximated by the vorticity
inside the pick- up
 coil area (the second case mentioned in the beginning of this
subsection).~\cite{hwang,tiesinga}
 This approximation hence does not take the contributions from $S_n$'s
outside the pick-up coil area into account and in this sense it corresponds to $d=0$.
 One conclusion from the present work is that for a more precise
comparison
 with experiments one should instead consider the opposite case of
large $d$.
 The results of our simulations are presented in Sec.~\ref{sec_results}. In the
following section we elucidate the relation between the flux-noise spectrum and
the complex impedance, or equivalently the complex conductivity.

\section{Flux Noise and Conductivity} \label{sec_theory}

In this section we consider for simplicity a 2D superconductor in the 
continuum limit so that, compared to the previous section, 
the limit $a\rightarrow 0$ is implied instead of $a=1$.
The magnetic flux associated with the area $d^2r$ around 
 ${\bf r}$ is then $n({\bf r})d^2r$ [see Eq.~(\ref{eq_nn})]. 
 In the Coulomb-gas analogy of vortices, $n({\bf r})$ is the charge 
density.~\cite{minnhagen-rev,jonsson,olsson}  
The charge density correlation function is given by 
 $c(r,t)=\langle n(r,t)n(0,0)\rangle$. We will first relate $c(r,t)$ 
 to the dielectric function $1/\hat{\epsilon}({\bf k},\omega)$ and the conductivity $\sigma(\omega)$ of the superconductor: 
The charge density correlation function    
$c(r,t)$ is related to the charge density response function $g(r,t)$ by  
\begin{equation}  
{\rm Im}[\hat{g}({\bf k},\omega)]=\frac{\omega}{2T^{CG}}\hat{c}({\bf k},\omega) ,  
\label{g}  
\end{equation}  
where $\hat{g}$ and $\hat{c}$ denote the Fourier transforms.  
The dielectric function 
$1/\hat{\epsilon}({\bf k},\omega)$ is given by the usual linear response relation~\cite{minnhagen-rev}  
\begin{equation}  
\frac{1}{\hat{\epsilon}({\bf k},\omega)}=1-\frac{2\pi}{k^2}\hat{g}({\bf k},\omega)   .
\label{eps}  
\end{equation}    
 
We define $1/\epsilon(\omega)\equiv 1/\hat{\epsilon}({\bf k}=0,\omega)$ 
which means that  
Eqs.~(\ref{g}) and (\ref{eps}) for ${\bf k}=0$ corresponds to 
Eqs.~(\ref{eq_Re_eps}) and (\ref{eq_Im_eps}) for the RSJ model.  
From Eqs.~(\ref{sigma}), (\ref{g}), and (\ref{eps}) we obtain a 
relation between the charge 
density correlations $c(r,t)$ and the real part of the conductivity 
$\sigma(\omega)$:  
\begin{equation}  
{\rm Re}[\sigma(\omega)]=-\frac{T}{2\pi
  T^{CG}\omega}{\rm Im}\left[\frac{1}{\epsilon(\omega)}\right] ,  
\label{sigma-eps}  
\end{equation}  
and  
\begin{equation}  
{\rm Im}\left[\frac{1}{\epsilon(\omega)}\right]=
-\frac{\pi\omega}{T^{CG}}\lim_{k\rightarrow 0}\frac{\hat{c}(k,\omega)}{k^2} .
\label{eps-c}  
\end{equation}  
  
Next we relate the charge density correlation function $c(r,t)$ to the 
flux-noise spectrum. From Eq.~(\ref{eq_full}) we have that the magnetic 
flux measured by the pick-up coil is 
\[ 
\Phi=\int_{{\rm coil}}B_z({\bf r})d^2r , 
\] 
where the integral is over the area covered by the coil. The magnetic 
field $B_z({\bf r})$ can be expressed as 
\[ 
B_z({\bf r})=\int f(|{\bf r}' -{\bf r}|,d)n({\bf r})d^2r'  , 
\] 
the $r'$-integration is over the whole 2D plane, and from 
Eqs.~(\ref{eq_nn}) and (\ref{eq_full}),  we have 
\begin{equation} 
f(r,d)=\frac{T}{T^{CG}}\frac{3d^2-(r^2+d^2)}{(r^2+d^2)^{\frac{5}{2}}} . 
  \label{eq_f} 
\end{equation} 
This means that the flux-noise spectrum 
$S(t)=\langle \Phi(t)\Phi(0)\rangle$ is given by 
\end{multicols} 
\noindent\rule{0.5\textwidth}{0.1ex}\rule{0.1ex}{2ex}\hfill 
\begin{equation} 
  S(t)=\int_{{\rm coil}}d^2r\int_{{\rm coil}}d^2r'\int d^2r''\int 
  d^2r'''f(|{\bf r}''-{\bf r}|,d)\langle n({\bf r},t)n({\bf r'},0)\rangle f(|{\bf r}'-{\bf r}'''|,d) 
  \label{eq_S} 
\end{equation} 
We can now use the convolution theorem and express $S(t)$ in terms of 
the Fourier transforms of $f(r,d)$ and $c(r,t)=\langle
n(r,t)n(0,0)\rangle$, i.e., 
\[ 
S(t)= \int_{{\rm coil}}d^2r\int_{{\rm
    coil}}d^2r'\int\frac{d^2k}{(2\pi)^2}e^{i{\bf k}
\cdot({\bf r}-{\bf r}')}|\hat{f}(k,d)|^2\hat{c}(k,t)  .
\] 
\hfill\raisebox{-1.9ex}{\rule{0.1ex}{2ex}}\rule{0.5\textwidth}{0.1ex} 
\begin{multicols}{2} 
\noindent
Taking the Fourier transform of $S(t)$ and using the
connection between $\hat{c}(k,\omega)$ and $1/\hat{\epsilon}({\bf k},\omega)$ given 
by Eqs.~(\ref{g}) and (\ref{eps}) yields
\begin{equation} 
S(\omega)=-\int_0^\infty dk\hat{F}(k,d) 
{\rm Im}\left[\frac{1}{\hat{\epsilon}(k,\omega)}\right]  
\label{eq_somega} 
\end{equation} 
where 
\begin{equation} 
\hat{F}(k,d)= \frac{2T^{CG}}{(2\pi)^2\omega} \int_{\rm coil}d^2r\int_{\rm coil}d^2r' 
e^{i{\bf k}\cdot({\bf r}-{\bf r}')}k^3|\hat{f}(k,d)|^2 , 
\label{eq_F} 
\end{equation} 
and $\hat{f}(k,d)$ is the Fourier transform of $f(r,d)$
which describes the spreading of the magnetic field. Within the
dipole approximation of Eq.~(\ref{eq_f}) $\hat{f}(k,d)$ is given
by
\begin{equation}
  \hat{f}(k,d)=\frac{T}{T^{CG}}2\pi ke^{-kd} . 
  \label{eq_kd}
\end{equation}    
The extreme case $d=0$ is outside the
dipole approximation and is given by
\begin{equation}
  \hat{f}(k)=\frac{T}{T^{CG}}.
  \label{point}
\end{equation}
The important point to note is that the flux-noise spectrum $S(\omega)$ is directly
related to the response function
 ${\rm Im}[1/\hat{\epsilon}(k,\omega)]$ through a function $\hat{F}(k,d)$
which contains all the
information of the spreading of the magnetic field above the
superconductor
as well as the geometry and position of the pick-up coil. 
 
In Ref.~\onlinecite{houlrik} it was suggested that the relation between the flux
noise $S(\omega)$
 and ${\rm Im}[1/\hat{\epsilon}(k,\omega)]$ could be further 
simplified to 
\[ 
S(\omega)\propto \frac{1}{\omega}\left|{\rm Im}
\left[\frac{1}{\hat{\epsilon}(0,\omega)}\right]\right| , 
\] 
or equivalently  
\begin{equation} 
S(\omega)\propto {\rm Re}[\sigma(\omega)] .
\label{eq_s-sigma} 
\end{equation} 
We will here show that for the typically experimental situation  
this proportionality between the conductivity  
and the flux noise is indeed valid. 
 
In order to establish this we assume for simplicity that the pick-up
coil is circular with radius $R$. In this case it is possible to
obtain an explicit
expression for $\hat{F}(k,d)$ in Eq.~(\ref{eq_F}), i.e.,
\begin{equation}
  \hat{F}(k,d)=\frac{2T^{CG}}{\omega}R^2 k|\hat{f}(k,d)|^2[J_1(kR)]^2 , 
  \label{eq_FR}
\end{equation}
where $J_1$ is the Bessel function of order one.
Let us first consider the limit in which the dipole approximation
Eq.~(\ref{eq_f}) is valid. This limit implies that the distance $d$ to the
pick-up coil is sufficiently large compared to the relevant microscopic lengths. For
a Josephson junction array the microscopic length is the lattice
constant $a$ which is typically of the order 1-10 $\mu m$ whereas for a
continuum superconductor it is the size of a vortex core typically
given by the Ginzburg-Landau coherence length $\xi$ and is of the order of
100-1000 {\AA} close to the transition. In addition $\Lambda$ has to be
much larger than $a$ for a Josephson junction array and $\xi$ for
a superconducting film. The typical size of $d$ is 100
$\mu m$.~\cite{ferrari,rogers,shaw,uppsala} 
So in practice the vortex dipole approximation may be expected to be
valid for superconducting films but the validity for Josephson 
junction arrays may be more questionable.
When the vortex dipole approximation is valid the noise spectrum 
is given by [compare Eqs.~(\ref{eq_somega}), (\ref{eq_kd}), and (\ref{eq_FR})] 
\end{multicols}
\noindent\rule{0.5\textwidth}{0.1ex}\rule{0.1ex}{2ex}\hfill
\begin{equation}
S_R(\omega)=-\frac{8\pi^2T^2R^2}{T^{CG}\omega} \int_0^\infty dk  k^3e^{-2kd}[J_1(kR)]^2 
{\rm Im}\left[\frac{1}{\hat{\epsilon}(k,\omega)}\right] .
\label{eq_somegaR}
\end{equation}
\hfill\raisebox{-1.9ex}{\rule{0.1ex}{2ex}}\rule{0.5\textwidth}{0.1ex}
\begin{multicols}{2}
\noindent
One notes that $k$-values much larger than $1/d$ will not contribute
to the integral in Eq.~(\ref{eq_somegaR}) because of the factor $e^{-2kd}$.  
This means that if $d$ is sufficiently large compared to the relevant microscopic
length, then ${\rm Im}[1/\hat{\epsilon}(k,\omega)]$ can be replaced
by ${\rm Im}[1/\hat{\epsilon}(0,\omega)]$, demonstrating that under
these conditions $S(\omega)$ is indeed proportional to ${\rm Im}[1/\omega\hat{\epsilon}(0,\omega)]$.
In the present case of a circular pick-up coil we explicitly find
\[
  S_R(\omega)=-\frac{C}{\omega} 
{\rm Im}\left[\frac{1}{\hat{\epsilon}(0,\omega)}\right]  , 
\] where the proportionality constant $C$ is given by (after changing
the integration variable to $x=kd$)
\[
C=\frac{8\pi^2T^2}{T^{CG}}\frac{R^2}{d^4}\int_0^\infty dx  x^3e^{-2x}[J_1(xR/d)]^2 .
\]  
In the limit $R/d\ll 1$ this reduces to
\begin{equation}
  C \approx \frac{T^2}{T^{CG}}\frac{R^4}{d^6}\frac{15\pi^2}{4}, 
  \label{eq_Rd1}
\end{equation}
and in the limit $R/d\gg 1$ to
\begin{equation}
 C \approx   \frac{T^2}{T^{CG}}\frac{2\pi R}{d^3} .
  \label{eq_Rd2}
\end{equation}
We conclude from Eqs.~(\ref{eq_Rd1}) and (\ref{eq_Rd2})
that the proportionality constant $C$ will always contain a
temperature-dependent factor $T^2/T^{CG}$ and a dependence on the
size of the coil. This coil-size dependence reflects the relative
magnitudes between the coil size and the distance from the sample:
When $R$ is much larger than $d$, the noise amplitude is proportional to
the perimeter of the coil $2\pi R$, and, when it is much smaller, it is
proportional to square of the coil area $R^4$. In typical
experiments $R$ is usually much larger than
$d$.~\cite{ferrari,rogers,shaw,uppsala}

In some previous simulations of the flux-noise spectrum, based on the $XY$ models, one
has approximated the flux-noise spectrum from the
fluctuation of the total vorticity for a finite area of the
model.~\cite{hwang,tiesinga} This implies two differences in relation to
the above dipole approximation: First of all it corresponds to $d=0$
and consequently to a constant $\hat{f}$ [see Eq.~(\ref{point})].  
Secondly it corresponds to changing the magnetic flux
defined by Eq.~(\ref{eq_nn}) to vorticity defined by Eq.~(\ref{eq_v1}).
Let us first consider the first change by itself: The case of a
circular area with radius $R$ then corresponds to the flux noise
[compare Eqs.~(\ref{eq_somega}), (\ref{point}), and (\ref{eq_FR})]
\begin{equation}
S_R(\omega)=-\frac{2T^2}{\omega T^{CG}} R^2 \int_0^\infty dk k[J_1(kR)]^2 
{\rm Im}\left[\frac{1}{\hat{\epsilon}(k,\omega)}\right] .
\label{eq_d0}
\end{equation}
This means that, in this case, the flux-noise spectrum depends on all the
$k$-values of ${\rm Im}[1/\hat{\epsilon}(k,\omega)]$ and is not
proportional to ${\rm Im}[1/\hat{\epsilon}(0,\omega)]$. For example
the leading large-$R$ dependence of Eq.~(\ref{eq_d0}) is
\[
S_R(\omega)\propto -\frac{R}{\omega}\int_0^\infty dk 
{\rm Im}\left[\frac{1}{\hat{\epsilon}(k,\omega)}\right] .
\]
So in this limit the flux noise is proportional to the perimeter of the pick-up
area just as for the large-$d$ case, but instead of singling out the
$k=0$-contribution all $k$-values contribute. The proportionality
between the flux noise $S(\omega)$ and
${\rm Im}[1/\omega \hat{\epsilon}(0,\omega)]\propto
{\rm Re}[\sigma(\omega)]$ can be tested by experiments since both
$S(\omega)$ and the conductivity $\sigma(\omega)$ can be
independently measured.~\cite{uppsala}

The change from magnetic flux
[defined by Eq.~(\ref{eq_nn})] to vorticity [defined by Eq.~(\ref{eq_v1})]
influences the flux-noise spectrum in an additional  significant way:
Now the crossing of a vortex over the perimeter of the pick-up area is
described as a
discrete $\pm 2\pi$-change of the total vorticity of the pick-up
area. The corresponding spectrum hence corresponds to a random walk of
discrete events over a sharp boundary. This results in a
$\omega^{-\frac{3}{2}}$-tail of the spectrum.~\cite{lax} However, this condition
  does not correspond to the experimental situation where
  the pick-up coil does not have a sharp boundary, is at a distance $d$
  from the sample, and, most importantly, the magnetic field
  from a vortex is spread out.

In the next section we present numerical results of the flux-noise
spectrum and its relation to the conductivity. 
 
\section{Simulation Results and Experimental Implications} \label{sec_results}

\subsection{Comparison with previous works} \label{sub_comp} 
\begin{figure}
\centerline{\epsfxsize=8cm \epsfbox{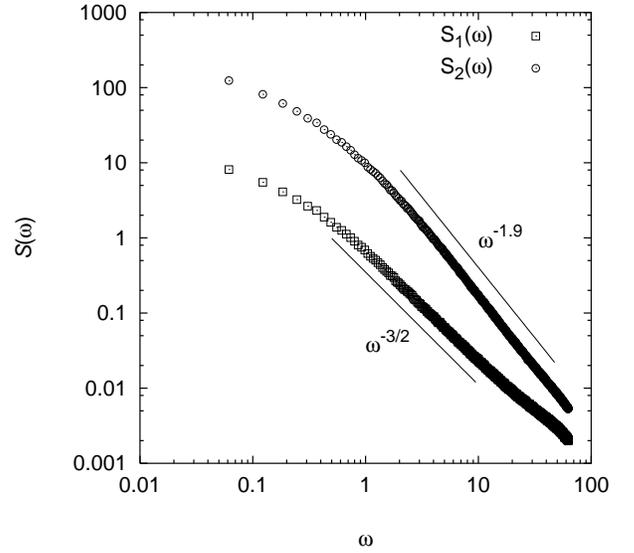}}
\vskip 0.5cm
\caption{Comparison between the vorticity-noise spectrum $S_1$ and the
  magnetic-flux-noise spectrum $S_2$ for $d=0$.
The data are for a $64 \times 64$ array at $T=1.1$ with a 
$32 \times 32$ coil size.
The vorticity spectrum has a $\omega^{-3/2}$ tail, whereas the
magnetic-flux spectrum is closer to $\omega^{-2}$.~\protect\cite{foot1} 
}
\label{fig_comp}
\end{figure}

We first relate our simulations results to earlier ones for the 
RSJ model.~\cite{hwang,tiesinga} 
These earlier simulations calculated the noise spectrum of 
the {\em vorticity} [see Eq.~(\ref{eq_v1})] over a fixed area of the 
systems (the $d=0$-case).~\cite{hwang,tiesinga} As explained in the 
previous section, this corresponds to discrete events over a sharp 
boundary and implying a $\omega^{-3/2}$-tail.~\cite{lax} Such a tail has indeed 
been found in Ref.~\onlinecite{hwang} and is also 
verified in  our simulations.~\cite{foot1} This is apparent from  Fig.~\ref{fig_comp} 
which displays our data for a $64\times 64$ array with a pick-up 
area of size  
$32\times 32$ at $T=1.1$; the lower data-set shows the {\em vorticity}-noise 
spectrum and the slope is $-3/2$. 
However, if we instead 
calculate the {\em magnetic-flux}-noise spectrum [see Eq.~(\ref{eq_nn})], 
then the crossing of magnetic flux over the perimeter 
is not a discrete event. This means that there is no obvious reason for a 
$\omega^{-3/2}$-tail and nor do we find any such tail in the 
simulations, as is also apparent from Fig.~\ref{fig_comp}. The exponent in Fig.~\ref{fig_comp} for 
this case is instead close to $-2$ as seen from the upper data-set in 
Fig.~\ref{fig_comp}.~\cite{foot2} 
\subsection{Flux-noise spectra with a finite distance  
between pick-up coil and array}\label{sub_d} 
As mentioned in Sec.~\ref{sec_theory} the typical experimental setup 
measures the {\em magnetic-flux}-noise spectrum with a finite distance 
$d$ between the array and the pick-up coil.~\cite{ferrari,rogers,shaw,uppsala} 
Furthermore the typical experimental setup corresponds to the large-$d$ 
limit where the noise spectrum $S(\omega)$ is proportional to the 
real part of the conductivity $\sigma(\omega)$ (see Sec.~\ref{sec_theory}). 
 
\begin{figure}
\centerline{\epsfxsize=10cm \epsfbox{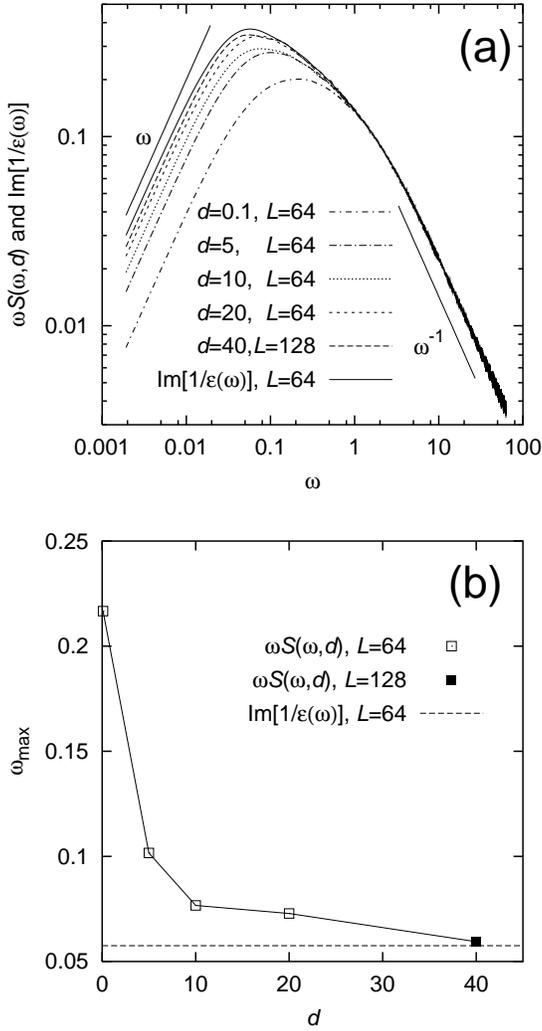}}
\vskip 0.5cm
\caption{(a) The dependence of the flux-noise spectrum $S(\omega, d)$ on
the distance $d$. The full drawn uppermost curve is the imaginary part
of the dielectric function $|{\rm Im}[1/\epsilon(\omega)]|$. The rest of the curves correspond
to $d = 0.1$, 5, 10, 20, and 40 (from bottom to top) plotted as
$\omega S(\omega, d)$ and the curves
are shifted in the vertical direction for better comparison.
The data are for a $64 \times 64$ array with a $32 \times 32$ pick-up coil at
$T=1.1$, except for  $d=40$ where a $128\times 128$ array was
necessary because plaquettes further away from the center contribute
in this case. As $d$ is increased $\omega S(\omega, d)$ approaches
$|{\rm Im}[1/\epsilon(\omega)]|$.  
(b) The frequency at the maxima for the curves in (a) are plotted versus the distance $d$.
As $d$ is increased, this frequency decreases and approaches the value
for $|{\rm Im}(1/\epsilon(\omega)|$. (The full line is a guide to the eye.)}
\label{fig_peak}
\end{figure}

In our simulations we investigate the flux-noise spectra as a 
function of the distance $d$ to the pick-up coil.  
Figure~\ref{fig_peak}(a) shows the flux-noise spectra 
calculated from Eqs.~(\ref{eq_Sw}), (\ref{eq_St}) with the magnetic flux 
given by Eqs.~(\ref{eq_Phi}), (\ref{eq_Vn}), and (\ref{eq_qn})
(see Sec.~\ref{sub_dipole} for 
details). The data sets are shown as $\omega S(\omega,d)$ against $\omega$ in 
a log-log plot. The vertical scale is adjusted in order to compare the 
shapes of the curves. One notices that the spectra for the different $d$'s 
all approach $\omega^{-1}$ for large $\omega$ and can be collapsed 
to a single curve in this large-$\omega$ region by a vertical 
adjustment. For small $\omega$ the curves becomes linear with $\omega$ 
which reflects a constant part (white noise) of $S(\omega,d)$.~\cite{shaw,uppsala,hwang,tiesinga} 
As $d$ 
increases the peak of the $\omega S(\omega,d)$ curves moves to the left 
and the peak height increases. 
The uppermost 
curve in Fig.~\ref{fig_peak}(a) is $|{\rm Im} [1/\epsilon(\omega)]|$ (full 
curve in the figure) and, as $d$ is increased, the flux-noise spectrum 
$\omega S(\omega,d)$ 
approaches this uppermost curve. This verifies that for large $d$ one has the 
simple connection $S(\omega)\propto |{\rm Im} [1/\omega\epsilon(\omega)]|$ as 
discussed in Sec.~\ref{sec_theory}. 
 
Figure~\ref{fig_peak}(b) shows that the characteristic frequency given 
by the peak position in Fig.~\ref{fig_peak}(a) decreases with increasing $d$. 
In the limit of large $d$ the characteristic frequency of 
$S(\omega,d)$ agrees with the characteristic frequency of 
$1/\epsilon(\omega)$. Thus in this limit both the shape  
and the characteristic frequency of $S(\omega,d)$ and 
${\rm Im}[1/\omega\epsilon(\omega)]$ are the same. This proportionality between
the flux-noise spectrum and complex conductivity can be tested experimentally and indeed 
seems to be borne out.~\cite{uppsala} 
 
\begin{figure}
\centerline{\epsfxsize=10cm \epsfbox{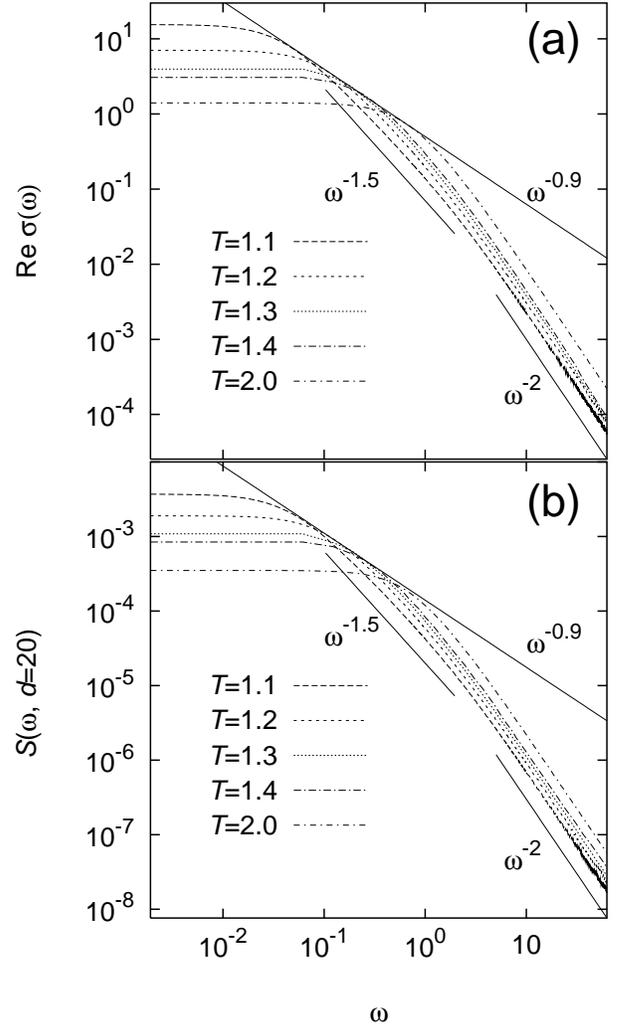}}
\vskip 0.5cm
\caption{(a) The real part of the conductivity ${\rm Re}[\sigma(\omega)]$
and (b) the flux-noise spectrum $S(\omega,d=20)$ 
at temperatures above the KT transition.
The curves for ${\rm Re}[\sigma(\omega)]$ and $S(\omega,d=20)$ have the
same shape; for small frequencies they have a very weak
$\omega$-dependence, for somewhat larger $\omega$ there is an
approximate $\omega^{-3/2}$-behavior, whereas for even larger
$\omega$ the behavior approaches $\omega^{-2}$. The curves in the
log-log plot have a common tangent with the slope $-0.9$.}
\label{fig_flux}
\end{figure}

The fact that the flux-noise spectrum in the large-$d$ limit is
proportional to real part of the conductivity means that the characteristic
features
of the conductivity are reflected in the flux-noise spectrum.
In case of a 2D superconductor the dynamical features of the
conductivity $\sigma(\omega)\propto -1/i\omega\epsilon(\omega)$, are well described by the response 
form~\cite{jonsson} 
\begin{equation} 
{\rm Re}\left[\frac{1}{\epsilon(\omega)}\right]- \frac{1}{\epsilon(0)}=\frac{1}{\tilde{\epsilon}}\frac{\omega}{\omega +\omega_0} , 
\label{reeps} 
\end{equation} 
and 
\begin{equation} 
{\rm Im}\left[\frac{1}{\epsilon(\omega)}\right]=-\frac{1}{\tilde{\epsilon}}\frac{2}{\pi} 
\frac{\omega\omega_0\ln \omega/\omega_0}{\omega^2 -\omega_0^2}  ,
\label{imeps} 
\end{equation} 
which catches the dynamics of vortex fluctuations in a region around
the KT transition. From Eq.~(\ref{imeps}) one
notices that the peak of $|{\rm Im}[1/\epsilon(\omega)]|$ occurs at the
characteristic frequency
$\omega_0$ and that the peak height is $1/\pi\tilde{\epsilon}$.
Above the KT transition $1/\tilde{\epsilon}$ increases only weakly
with increasing temperature and approaches unity for somewhat higher
temperatures.~\cite{jonsson} For the flux-noise spectrum this means that
$S(\omega_0)\propto T^2/T^{CG}\tilde{\epsilon}\omega_0$.
Now $T/T^{CG}\propto \rho_0(T)$ where $\rho_0(T)$ is the bare
superfluid density which decreases slightly with
temperature~\cite{minnhagen-rev} whereas $T$ increases so that
also $T^2/T^{CG}$ depends only weakly on $T$. This means that to good
approximation the flux-noise spectrum for different temperatures
above the KT transition should have a common tangent$\propto
1/\omega$ which goes through all the points $S(\omega_0(T),T)$. This
means that the common tangent in a log-log plot has the slope $-1$. 
This feature is illustrated in Fig.~\ref{fig_flux} which demonstrates
the existence of a common tangent with the slope close to $-1$ both
directly for $S(\omega,d)$ and for
${\rm Re}[\sigma(\omega)]\propto -{\rm Im}[1/\omega\epsilon(\omega)]$.
The existence of a common tangent for the flux-noise spectra
with the slope $-1$ can be readily tested in experiments and seems
to be well borne out.~\cite{uppsala,minnhagentriest} 
 
One should, however, notice that the argument for a common tangent
with slope $-1$ does not single out the response form given by
Eqs.~(\ref{reeps}) and (\ref{imeps}). In fact the reasoning is also
valid in a region above the KT transition  
where the response of the vortex fluctuations is given by the conventional Drude response form 
\begin{equation} 
{\rm Re}\left[\frac{1}{\epsilon(\omega)}\right]=\frac{1}{\tilde{\epsilon}}\frac{\omega^2}{\omega^2+\omega_0^2} , 
\label{druder} 
\end{equation} 
and 
\begin{equation} 
{\rm Im}\left[\frac{1}{\epsilon(\omega)}\right]=-\frac{1}{\tilde{\epsilon}}\frac{\omega\omega_0}{\omega^2+\omega_0^2} .
\label{drudei} 
\end{equation} 
which gives the peak height $1/2\tilde{\epsilon}$. 
One expects that the response form Eqs.~(\ref{reeps}) and (\ref{imeps})  
describes the response from the vortex pairs in a region just above the  
KT transition whereas the conventional Drude response is obtained for
higher temperatures where the response is dominated by free
vortices.~\cite{jonsson} How wide these regions are depend on
the details: for a real thin superconductor the vortex-pair response
seems to dominate in a wide region.~\cite{jonsson} However, for the
2D RSJ model on a square lattice, which we are using here, the
vortex-pair dominated region above the KT transition is narrow
and the Drude response dominates in a broader region
above the KT transition. Thus the data shown in Fig.~\ref{fig_flux} are
predominantly Drude-like. A practical way of determining which response type
is at hand is to measure the complex impedance and determine the peak
ratio [i.e., the ratio ${\rm Re}(\sigma)/{\rm Im}(\sigma)$
at the peak position $\omega_0$ of ${\rm Re}(\sigma/\omega)$];
for the vortex pair dominated response, Eqs.~(\ref{reeps}) and
(\ref{imeps}),
this ratio is $2/\pi\approx 0.64$ and for free vortex Drude response, 
Eqs.~(\ref{druder}) and (\ref{drudei}), it is unity.~\cite{jonsson} 

\begin{figure}
\centerline{\epsfxsize=9cm \epsfbox{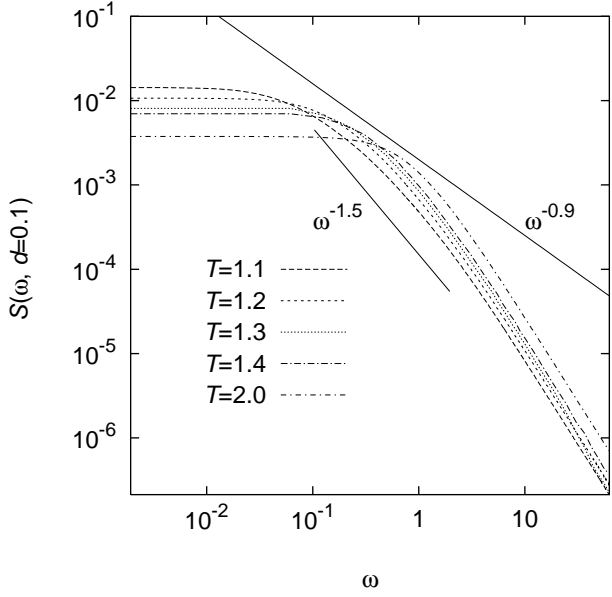}}
\vskip 0.5cm
\caption{Flux-noise spectrum in the small-$d$ limit
at temperatures above the KT transition [same as 
Fig.~\protect\ref{fig_flux}(b) but with $d=0.1$ instead of $d=20$]. 
In the case of small $d$ we find neither 
a common tangent nor an appreciable range of $\omega^{-1.5}$-behavior.}
\label{fig_d0.1}
\end{figure}

The essential point here is that, because the flux-noise spectrum
for a large $d$ is proportional to ${\rm Re}[\sigma(\omega)]$,
the noise spectra at a sequence of temperatures above the
KT transition should in a log-log plot have a common tangent with slope $-1$.
We can substantiate this claim further by simulating the noise spectra
for a small $d$ where the flux-noise spectrum is {\em not}
proportional to ${\rm Re}[\sigma(\omega)]$. In this case there is no particular reason
for a common tangent with any slope and, as apparent from the
simulation results in Fig.~\ref{fig_d0.1}, no such common tangent can be fitted to the data. 
 
One may also notice from Fig.~\ref{fig_flux} that both $S(\omega,d)$
and ${\rm Re}[\sigma(\omega)]$ have intermediate regions with
$\omega^{-1.5}$ followed by a $\omega^{-2}$-tail for even larger $\omega$.
However, such an intermediate-$\omega^{-1.5}$ region appears to be
less discernible for
the small-$d$ case when $S(\omega,d)$ is not proportional to
${\rm Re}[\sigma(\omega)]$,
as is apparent from Fig.~\ref{fig_d0.1}. 
 
\subsection{Flux-noise spectrum below KT transition}\label{sub_below} 
\begin{figure}
\centerline{\epsfxsize=10cm \epsfbox{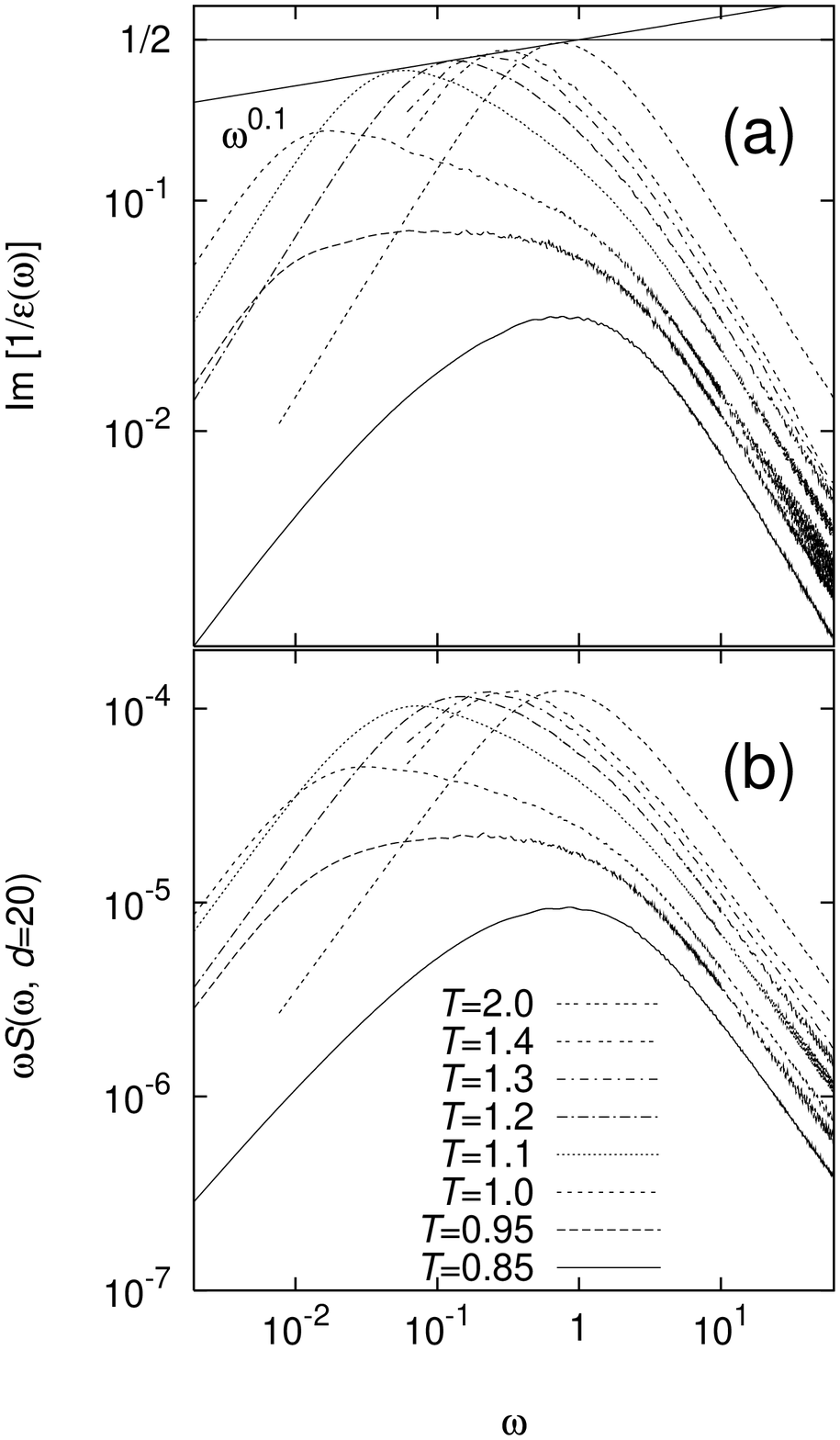}}
\vskip 0.5cm
\caption{(a) The imaginary part of the dielectric function, $|{\rm Im}[1/\epsilon(\omega)]|$,
and (b) the flux-noise spectrum multiplied by the frequency, $\omega S(\omega,d=20)$, 
at temperatures above and below the KT transition.  This again
illustrates that both
quantities behave in the same way. As the temperature is increased far above
the KT transition, the maximum of $|{\rm Im}[1/\epsilon(\omega)]|$ approaches
the limit value $1/2$ (this corresponds to the Drude limit with $\tilde\epsilon = 1$) 
as denoted by the horizontal line in (a). The curves seem to
develop a plateau as the KT transition is approached from above, as
is suggested by the $T=0.95$-curves. As the temperature drops below
the KT transition, the amplitude of the flux noise rapidly decreases
whereas the characteristic frequency increases, as is illustrated by
the curves at $T=0.85$. 
}
\label{fig_all}
\end{figure}

Next we investigate what happens as the temperature is decreased towards the KT transition and below. 
Figure~\ref{fig_all} shows data for ${\rm Im}[1/\epsilon(\omega)]$ and $\omega S(\omega,d=20)$ 
over a wider range of temperatures (the data for $T\geq 1.1$ are the same as in Fig.~\ref{fig_flux}). 
Again one observes that both quantities behave in precisely the same
way over the whole temperature range, verifying that they are indeed
proportional to very good approximation. Next one observes that the
characteristic frequency $\omega_0$ (the frequency of the peak
position) decreases as the KT transition is approached from above.
This suggests a critical slowing down at the KT transition to
$\omega_0=0$.~\cite{hwang,jonsson} As the temperature passes through
the transition the characteristic frequency starts to increase again
(the full curves in Fig.~\ref{fig_all} are below the KT transition).~\cite{hwang,jonsson}  
 
The argument for a common $1/\omega$-tangent for the flux-noise
spectra above the KT transition is related to the fact that the peak height for
$|{\rm Im}[1/\epsilon(\omega)]$ is $1/\pi
\tilde{\epsilon}$ ($1/2\tilde{\epsilon}$)
for the vortex pair response, Eqs.~(\ref{reeps}) and
(\ref{imeps}) [free vortex Drude response, Eqs.~(\ref{druder}) and
(\ref{drudei})] together with the fact that $1/\tilde{\epsilon}$
only increases very weakly with $T$ above 
the KT transition and approaches unity for somewhat higher $T$.
Figure~\ref{fig_all} shows this weak increase in a region above the
KT transition; the Drude value 1/2 is approached roughly like
$\omega^{0.1}$
which explains the discrepancy between exponent $1$ and the value $0.9$ 
found in Fig.~\ref{fig_flux}. Thus the existence of a common tangent $1/\omega$
hinges on the weakness
of the temperature dependent factor $T^2/T^{CG}(T)\tilde{\epsilon}(T)$
which in turn depends somewhat on the details of the system. However,
since the temperature dependence of $\omega_0$ is dramatic just above
the KT transition,
the common tangent should to good approximation exist at least
in a limited region above the KT transition.  
 
Below the KT transition there are no free vortices and the response is
given by the vortex pairs Eqs.~(\ref{reeps}) and (\ref{imeps}).~\cite{beom} 
However, in this case the factor $1/\tilde{\epsilon}\approx
1-1/\epsilon(\omega=0)$
decreases rapidly towards zero as the temperature is decreased
below the KT transition.~\cite{jonsson} This means that while
the characteristic frequency rapidly increases, as the temperature is
decreased below the KT transition, the amplitude of the flux noise,
which is proportional to $1/\tilde{\epsilon}\approx
1-1/\epsilon(\omega=0)$, rapidly decreases.
The KT transition is at $T\approx 0.9$ and already at $T=0.85$ the amplitude 
of $\omega S(\omega,d)$ has dropped dramatically compared to the almost constant 
amplitude above the KT transition, as is apparent from Fig.~\ref{fig_all}(b).       
 
Finally, there is in Fig.~\ref{fig_all}(b) an indication that the curves
develop a plateau as the KT transition is approached from above
(compare the curve for $T=0.95$). Such a plateau would suggest that
$S(\omega,d)$ is proportional to $1/\omega$ in an intermediate region
just above the KT transition. The same development of a plateau
can be anticipated in Fig.~\ref{fig_all}(a) for $|{\rm Im}[1/\epsilon(\omega)]|$ and
has also been found for the $XY$ model with the time-dependent
Ginzburg-Landau (TDGL) dynamics.~\cite{jonsson,minnhagentriest}
 
\section{Concluding Remarks} \label{sec_conc}

In the present paper we have explored the fact that the typical
experimental setup for measuring the magnetic flux noise for a
superconductor, or JJA, corresponds to the case when the distance $d$ to
the pick-up coil is much larger than the relevant microscopic
lengths. In this limit the flux-noise spectrum and the real part of the
conductivity are proportional. This proportionality seems first to have
been anticipated in Ref.~\onlinecite{houlrik} and has also been
experimentally verified.~\cite{uppsala} In this paper we have studied this
connection in some more detail.

We have also demonstrated that both the shape
and the characteristic frequency of the spectrum depend  on the
distance $d$ to the pick-up coil. This means that no detailed
conclusions can be drawn from
simulations which presumes $d=0$. Furthermore, there is a qualitative
difference between the {\em vorticity}-noise spectrum, which corresponds to
discrete events over a sharp boundary, and the {\em magnetic-flux}-noise
spectrum which corresponds to spread-out objects over a boundary.
The experimental situation corresponds to a {\em spread-out magnetic flux}
and a pick-up coil at a large distance $d$,
which is very different
from some earlier simulations which calculated the
{\em vorticity}-noise spectrum for $d=0$.~\cite{hwang,tiesinga}
Nevertheless the {\em vorticity}-noise spectrum for $d=0$ has a $\omega^{-3/2}$ tail for
higher frequencies which seems to match the experimental
results,~\cite{rogers,uppsala} whereas the {\em magnetic-flux}-noise spectra for
$d=0$ does not have such a tail. In accordance with the present
simulations, we suggest that the resolution of this
dichotomy is that in the large-$d$ limit the magnetic-flux-noise spectrum
does have an {\em intermediate} region with a $w^{-3/2}$ behavior and
that it is this intermediate region which is seen in the experiments.

The proportionality between the magnetic-flux-noise spectrum and the
real part of the conductivity implies that the noise spectra for a
sequence of temperatures just above the KT transition should in a
log-log plot have a common tangent with the slope $-1$. The existence
of such a common tangent has also be verified in
experiments,~\cite{uppsala} as well as in the present simulations.     
We also explicitly demonstrated through our simulations that for small $d$
there is no such common tangent.

The existence of a common tangent is by itself not necessarily
conclusive. For example, the experimental data for the JJA's in Ref.~\onlinecite{shaw}
correspond to the large-$d$ case and the data have indeed a 
common tangent with slope $-1$. However, the spectra at a fixed
temperature seem to have a $1/\omega$ behavior over a very large
region, which differ markedly from the spectra obtained in our simulations for
the RSJ model.  

The present simulations also suggest that immediately above the KT transition
there should be a very small temperature region where the noise spectrum has an
intermediate interval with a $1/\omega$ behavior.~\cite{minnhagentriest}
It has been suggested that the data in Ref.~\onlinecite{shaw} might
perhaps be related to this temperature region closest to the
transition.~\cite{minnhagentriest}
However, at the
moment there seem to be no accepted explanation for the
$1/\omega$-behavior found in Ref.~\onlinecite{shaw}.~\cite{hwang}

Finally, we showed that the amplitude of the flux-noise spectrum
drops dramatically as the temperature is decreased below the
KT transition, and that at the same time the characteristic frequency
increases. It should also be possible to observe this effect in experiments.

\acknowledgments
The authors are grateful to P.~Svedlindh and \"{O}.~Festin for
discussions of their experimental data and to D. Bormann for
discussions of the theory. This work was supported by the Swedish
Natural Research Council through contract FU 04040-332.

\narrowtext

\end{multicols}

\begin{thebibliography}{99}
\bibitem{foot-nr} Note that this only gives a measure of the flux. The magnetic field
associated with this flux is spread out; it is strongest in the vortex core 
region and decays on the length scale $\Lambda$.
\bibitem{foot-m} Note that the circulation of the normal current around a plaquette
vanishes since it is proportional to $\sum_p(\dot\theta_i-\dot\theta_j)=0$.
\bibitem{foot1} In this respect our results and the ones in
  Ref.~\onlinecite{hwang} agree completely but differ slightly from
  Ref.~\onlinecite{tiesinga} where  an exponent
  somewhat smaller than $3/2$ was obtained for the same case
  (i.e., $\approx 1.2$ at $T \gtrsim 1.1$).
\bibitem{foot2} In Fig.~\ref{fig_comp} there appears to be a slight upturn of the
  spectra at the largest frequencies. As in Ref.~\onlinecite{hwang},
  this is an artifact of the fast Fourier transformation used in the calculations.
\end{thebibliography}
\end{document}